\documentclass[conference]{IEEEtran}
\IEEEoverridecommandlockouts
\usepackage{listings, xcolor}

\definecolor{verylightgray}{rgb}{.97,.97,.97}

\lstdefinelanguage{Solidity}{
	keywords=[1]{anonymous, assembly, assert, balance, break, call, callcode, case, catch, class, constant, continue, constructor, contract, debugger, default, delegatecall, delete, do, else, emit, event, experimental, export, external, false, finally, for, function, gas, if, implements, import, in, indexed, instanceof, interface, internal, is, length, library, log0, log1, log2, log3, log4, memory, modifier, new, payable, pragma, private, protected, public, pure, push, require, return, returns, revert, selfdestruct, send, solidity, storage, struct, suicide, super, switch, then, this, throw, transfer, true, try, typeof, using, value, view, while, with, addmod, ecrecover, keccak256, mulmod, ripemd160, sha256, sha3}, % generic keywords including crypto operations
	keywordstyle=[1]\color{blue}\bfseries,
	keywords=[2]{address, bool, byte, bytes, bytes1, bytes2, bytes3, bytes4, bytes5, bytes6, bytes7, bytes8, bytes9, bytes10, bytes11, bytes12, bytes13, bytes14, bytes15, bytes16, bytes17, bytes18, bytes19, bytes20, bytes21, bytes22, bytes23, bytes24, bytes25, bytes26, bytes27, bytes28, bytes29, bytes30, bytes31, bytes32, enum, int, int8, int16, int24, int32, int40, int48, int56, int64, int72, int80, int88, int96, int104, int112, int120, int128, int136, int144, int152, int160, int168, int176, int184, int192, int200, int208, int216, int224, int232, int240, int248, int256, mapping, string, uint, uint8, uint16, uint24, uint32, uint40, uint48, uint56, uint64, uint72, uint80, uint88, uint96, uint104, uint112, uint120, uint128, uint136, uint144, uint152, uint160, uint168, uint176, uint184, uint192, uint200, uint208, uint216, uint224, uint232, uint240, uint248, uint256, var, void, ether, finney, szabo, wei, days, hours, minutes, seconds, weeks, years},	% types; money and time units
	keywordstyle=[2]\color{teal}\bfseries,
	keywords=[3]{block, blockhash, coinbase, difficulty, gaslimit, number, timestamp, msg, data, gas, sender, sig, value, now, tx, gasprice, origin},	% environment variables
	keywordstyle=[3]\color{violet}\bfseries,
	identifierstyle=\color{black},
	sensitive=false,
	comment=[l]{//},
	morecomment=[s]{/*}{*/},
	commentstyle=\color{gray}\ttfamily,
	stringstyle=\color{red}\ttfamily,
	morestring=[b]',
	morestring=[b]"
}

\usepackage{cite}
\usepackage{amsmath,amssymb,amsfonts}
\usepackage{algorithm} %replace with algortihmic
\usepackage{algpseudocode} %replace with algortihmic
\usepackage{graphicx}
\usepackage{textcomp}
\usepackage{xcolor}
\usepackage{comment}
\usepackage{pifont}
\usepackage{framed}
\usepackage{pgf-umlsd}
\usepackage{listings}
\usepackage{dblfloatfix}
\usepackage{threeparttable}
\usepackage{listings}
\def\BibTeX{{\rm B\kern-.05em{\sc i\kern-.025em b}\kern-.08em
    T\kern-.1667em\lower.7ex\hbox{E}\kern-.125emX}}
\newcommand{\cmark}{\ding{51}}%
\newcommand{\xmark}{\ding{55}}%

\newcommand{\minuseq}{\mathrel{{-}{=}}}

\errorcontextlines\maxdimen

\makeatletter
\newcommand*{\algrule}[1][\algorithmicindent]{\makebox[#1][l]{\hspace*{.5em}\vrule height .75\baselineskip depth .25\baselineskip}}%

\newcount\ALG@printindent@tempcnta
\def\ALG@printindent{%
    \ifnum \theALG@nested>0% is there anything to print
        \ifx\ALG@text\ALG@x@notext% is this an end group without any text?
            \addvspace{-3pt}% FUDGE for cases where no text is shown, to make the rules line up
        \else
            \unskip
            \ALG@printindent@tempcnta=1
            \loop
                \algrule[\csname ALG@ind@\the\ALG@printindent@tempcnta\endcsname]%
                \advance \ALG@printindent@tempcnta 1
            \ifnum \ALG@printindent@tempcnta<\numexpr\theALG@nested+1\relax% can't do <=, so add one to RHS and use < instead
            \repeat
        \fi
    \fi
    }%
\usepackage{etoolbox}
\patchcmd{\ALG@doentity}{\noindent\hskip\ALG@tlm}{\ALG@printindent}{}{\errmessage{failed to patch}}
\makeatother
\algnewcommand{\IfThenElse}[3]{% \IfThenElse{<if>}{<then>}{<else>}
  \algorithmicif\ #1\ \algorithmicthen\ #2\ \algorithmicelse\ #3}

\algrenewcommand{\algorithmicrequire}{\textbf{Input:}}
\algrenewcommand{\algorithmicensure}{\textbf{Output:}}

\begin{document}

\title{A Decentralized Framework with Dynamic and Event-Driven Container Orchestration at the Edge}

\author{\IEEEauthorblockN{Umut Can Özyar}
\IEEEauthorblockA{\textit{Computer Engineering Department} \\
\textit{Bogazici University}\\
Istanbul, Turkey \\
umut.ozyar@boun.edu.tr}
\and
\IEEEauthorblockN{Arda Yurdakul}
\IEEEauthorblockA{\textit{Computer Engineering Department} \\
\textit{Bogazici University}\\
Istanbul, Turkey \\
yurdakul@boun.edu.tr}
}

\maketitle

\begin{abstract}
% Eski Abstract (Duzeltmelerden onceki)
% Virtualization provides an abstraction layer for the Internet of Things technology to tackle the heterogeneity of the edge networks. It enables the deployment of an application on devices with different architectures to achieve uniformity. This study lays down the fundamentals of a framework for dynamic and event-driven orchestration towards a fully decentralized edge. It provides a blockchain-based delivery platform for containerized applications registered with their resource requirements through a registry on a distributed file system, namely InterPlanetary File System (IPFS). The decentralized resource manager running on the metrics scraped from the host and the virtualization platform, i.e., Docker in our implementation, dynamically optimizes the resources allocated to each container. The framework ensures that variable workloads of a heterogeneous environment can co-exist on multiple edge devices. An event-driven architecture is built over a lightweight messaging protocol, MQTT, capitalizing on the asynchronous and distributed nature of the publish/subscribe pattern to achieve a truly distributed system.
% Duzeltme #1
% Please show your research motivation after the background in the abstract.
% oneri umut, intro 2. paragrafi ozetle ve 3. paragrafin ozeti
Virtualization provides an abstraction layer for the Internet of Things technology to tackle the heterogeneity of the edge networks. Deploying virtualized applications on different architectures requires autonomous scaling and load balancing while ensuring their authenticity. A decentralized end-to-end solution is necessary for applications with variable workloads to co-exist on a heterogeneous environment with multiple edge devices. Hence, this study lays down the fundamentals of a framework for dynamic and event-driven orchestration towards a fully decentralized edge. It provides a blockchain-based delivery platform for containerized applications registered with their resource requirements through a registry on a distributed file system. The decentralized resource manager running on the metrics scraped from the host and the virtualization platform, i.e., Docker in our implementation, dynamically optimizes the resources allocated to each container. An event-driven architecture is built over a lightweight messaging protocol, MQTT, capitalizing on the asynchronous and distributed nature of the publish/subscribe pattern to achieve a truly distributed system.
\end{abstract}


\section{Introduction}
The term “Internet of Things (IoT)” describes a sophisticated system of heterogeneous devices, dynamic environments, and complex sub-systems \cite{gubbi2013internet}. The quality of IoT services delivered to the end-users is a major concern of the application developers. Edge computing has been proposed to improve user experience by moving services, processing, and actionable insights closer to the user. However, edge devices are also heterogeneous in terms of processing power, memory amount, and operating system. Hence, virtualization schemes, such as containerization, widely used in computer systems have recently stepped into the domain of edge computing \cite{tao2019survey}. As containers can be scaled according to the available resources of the host system, application developers can present an IoT solution in a container that should ideally execute on almost every edge device \cite{dolui_towards_2018}.

Even though the usage of containers is a breath-taking solution for IoT application developers, it bears its own problems that have to be resolved for a seamless user experience. 
Firstly, as edge devices are implemented with different types of hardware, container scaling should be done at the edge autonomously. Since each application can have a unique workload during runtime, dynamic load balancing has to be taken into account to maximize resource utilization while scaling the containers. 
Secondly, the authenticity of the container should be ensured prior to its deployment or upgrade, because malicious containers can overwhelm the host edge device which may be running on scarce resources. 
% Duzeltme #2
% The sentence “To resolve this issue, autonomous deployment of containers to neighboring devices should be considered” is unclear to read. You have mentioned three flaws before “resolve this issue”, but what does the “issue” refer to? Try “issues” or other presentations to express your point clearly
% Eski kullanim:
% Finally, the user should not experience an irresponsive or latent service at the edge. To resolve this issue, autonomous deployment of containers to neighboring devices should be considered. 
% Option 0:
Finally, irresponsive or latent services at the edge should be avoided to improve user experience through autonomous deployment of containers to neighboring devices.
% Option 1:
% Finally, the user should not experience an irresponsive or latent service at the edge which can be achieved by autonomous deployment of containers to neighboring devices.
% Option 2:
% Finally, the user should not experience an irresponsive or latent service at the edge. The user experience can be improved by autonomous deployment of containers to neighboring devices.

% Duzeltme #3
% Please present your contribution in the Introduction.
% our contributions can be summarizes as follows 1,2 3
% our contributions kullanabilir miyiz dusun
% olmuyor our contributions :(
In this study, we propose a decentralized framework for autonomous deployment and scaling of containerized applications on resource-constrained edge devices. The secure delivery of containerized applications is enabled by a smart contract that holds container registries on the blockchain. The container images are stored in a decentralized file system, namely, IPFS \cite{benet_ipfs_2014}. The use of blockchain and IPFS provides the ubiquity of the applications for a seamless user experience. Once the user registers to a service or employs an IoT device, the container deployment is done autonomously based on available on-device resources. An event-driven decentralized resource manager is designed for this purpose. It analyzes running services and forecasts future requirements. Based on its findings, it scales  containers running on the device. The decision about the new container depends on where the edge device is deployed. If there exist multiple devices at the edge, their resource managers talk, and the device with maximum abundant resources employs the container. Otherwise, only on-device resources are used in giving the decision. Our study is unique in the sense that it provides an end-to-end solution between the application release and the user experience.

The rest of the paper is organized as follows: The next section presents related studies in the literature. Section III introduces the proposed framework while explaining design choices and incorporated technologies. Section IV analyzes the behavior of the framework under different experimental setups. The final section concludes the work.

\section{Related Works}

In the recent literature, there exist several studies on edge container orchestration frameworks. An overview is presented in Table \ref{table:literature_comparison}.  

As mentioned in the previous section, the heterogeneous nature of IoT end devices requires scaling. Hence, autonomous scaling (C1) of the containers has to be done at the edge for a seamless user experience. If supported by the system, the options are horizontal (H) or vertical (V). Horizontal scaling refers to changing the number of containers of an application to meet the varying loads. Performance on a single-CPU single-threaded edge device may be degraded in horizontally-scaled containers because having multiple instances of the same application will have to execute sequentially. Thus, \cite{xiong_extend_2018,muralidharan2019monitoring,baresi_paps_2019,yang_kubehice_2021,ajayi_beca_2021-1,subramanya2021centralized} explore horizontal scaling across all available devices in their networks. Vertical scaling is the adaptation of the resources of an existing container. Containerization engines such as Docker provide all necessary tools for vertical scaling, making it inherently more straightforward to configure than horizontal scaling. Scaling type is tightly correlated with software heterogeneity (C3), such as accommodation of different kinds of applications like microservices, batch jobs, and streaming applications on the same device \cite{xiong_extend_2018,muralidharan2019monitoring,yang_kubehice_2021,struhar_react_2021,goethals_fledge_2020,pires_distributed_2021,cui_blockchain-based_2021}. Since our target is the utilization of software heterogeneity on resource-constrained edge devices, vertical scaling is preferred. 

Decision-taking for autonomous scaling can be either reactive or predictive. Reactive methods only consider the current state. Predictive methods (C2) consider the past and future states of the system. 
%todo subramanya2021centralized  hem horizontal hem de vertical scaling yapıyor ama ML algorıtmalarını edge calıstırmakta sorun yasıyor
In \cite{subramanya2021centralized} predictive methods such as machine learning are proposed for horizontal scaling. When vertical scaling is considered, predictive solutions provide a smoother user experience and better decision performance than reactive ones. Our framework relies on statistical models for time-series forecasting as a reliable method that doesn't depend on large amounts of training data. Auto-scaling decisions are based on optimization metrics (C6). The two main categories are application (A) and system (S) metrics. Application metrics are driven by the application requirements such as response time and error rate \cite{baresi_paps_2019,ajayi_beca_2021-1,yang_kubehice_2021,subramanya2021centralized}. System metrics, such as CPU and memory, are harnessed from the host. Application metrics fail to offer a viable solution when it has to execute on different types of hardware. Hence, our framework utilizes system metrics so that the applications can run on all types of hardware. Frameworks of \cite{dolui_towards_2018,xiong_extend_2018,muralidharan2019monitoring,goethals_fledge_2020,subramanya2021centralized,cicconetti_decentralized_2021} focus on the resource-constrained devices (C4) as we do, while the rest are deployed on more powerful devices on the edge. This makes on-device orchestration (C5) a challenging issue  since it also consumes system resources \cite{dolui_towards_2018,muralidharan2019monitoring,baresi_paps_2019,ajayi_beca_2021-1,pires_distributed_2021,subramanya2021centralized,cicconetti_decentralized_2021}. In this study, on-device orchestration is also adopted as it supports decentralization.  %However, \cite{subramanya2021centralized} risks slowing down processes by running machine learning algorithms on the same device.

%This variance depends on various aspects such as availability, load-balancing, scaling, and guaranteed resource allocation (C5). Our framework guarantees all these four aspects.  

Edge computing paradigm inherently supports the decentralization of every IoT service \cite{pires_distributed_2021}. However, the studies in the literature suffer from centralized application registries throughout the whole delivery process. A centralized registry gives the authority to set download rates, storage limits, or pricing decisions to a single party, granting them unfair advantages as well as introducing a single point of failure. Traditional decentralized registry solutions may suffer from some security issues such as access control, immutability, and authenticity of applications. To cope with this limitation, our decentralized delivery process leverages blockchain technology and a tamper-proof distributed file system to provide proof of ownership on the smart contracts (C8). Decentralized communication between edge devices is also essential. Event-driven communication (C7) using the publish/subscribe pattern is adopted in many studies \cite{dolui_towards_2018,xiong_extend_2018,muralidharan2019monitoring,cui_blockchain-based_2021,ajayi_beca_2021-1} as its resilience and robustness make it a popular choice on the edge for building scalable platforms. It is also strategic for decentralized load balancing (C9) across a cluster of edge devices autonomously \cite{cicconetti_decentralized_2021,baresi_paps_2019,pires_distributed_2021} to distribute the load on other edge devices. Since these works assume trustless setups, a consensus mechanism is implemented. Since we consider secured resource-constrained devices and blockchain is used for secure application delivery, a two-step broadcast followed by an approval-by-silence mechanism is proposed in this work.

%Our framework with a decentralized application registry offers a decentralized application delivery process (C8), unlike the rest of the reviewed works. Whether it is Docker's own registry called Docker Hub or a privately hosted one, they rely on centralized registries \cite{noauthor_docker_nodate}. Hence, they stick to centralized solutions throughout the whole delivery process. However, centralization gives the authority to set download rates, storage limits, or pricing decisions to a single party, granting them unfair advantages as well as introducing a single point of failure. Ours and frameworks in \cite{ajayi_beca_2021-1,xiong_extend_2018,muralidharan2019monitoring,cui_blockchain-based_2021} rely on event-driven communication (C7), as its adoption fits the distributed architecture of our framework. These frameworks benefit from many-to-many communication using the publish/subscribe pattern. Its resilience and robustness make it a popular choice on the edge for building scalable platforms. It is also strategic for decentralized load balancing (C9) across a cluster of edge devices autonomously, which is offered by \cite{baresi_paps_2019,cicconetti_decentralized_2021,pires_distributed_2021} as well as ours. In these frameworks, the host device selection for a container depends on a decentralized process like a consensus mechanism instead of this decision being made by a central authority. Decentralization ensures that this decision is vendor-agnostic and that all applications are treated fairly during orchestration.

\begin{table}
\begin{threeparttable}
\caption{Overview of Edge Orchestration Frameworks.}
\begin{tabular}{|c|c|c|c|c|c|c|c|c|c|}
\hline
\textbf{}&\multicolumn{9}{|c|}{\textbf{Criteria\textsuperscript{1}}} \\
\cline{2-10} 
\textbf{Authors} & \textbf{\textit{C1}}& \textbf{\textit{C2}}& \textbf{\textit{C3}}& \textbf{\textit{C4}}& \textbf{\textit{C5}}& \textbf{\textit{C6}}& \textbf{\textit{C7}}& \textbf{\textit{C8}}& \textbf{\textit{C9}} \\
\hline

% 3 Towards Multi-Container Deployment on IoT Gateways
\cite{dolui_towards_2018} & \xmark & \xmark & \xmark & \cmark & \cmark & \xmark & \cmark & \xmark & \xmark      \\
% 5 Extend Cloud to Edge with KubeEdge
\cite{xiong_extend_2018} & H & \xmark & \cmark & \cmark & \xmark & \xmark & \cmark & \xmark & \xmark      \\
% 6 Monitoring and managing iot applications in smart cities using kubernetes
\cite{muralidharan2019monitoring} & H     & \xmark & \cmark & \cmark & \cmark & S & \cmark & \xmark & \xmark      \\
% 7 PAPS- A Framework for Decentralized Self-management at the Edge
\cite{baresi_paps_2019} & H      & \xmark & \xmark & \xmark & \cmark & A & \xmark & \xmark & \cmark      \\
% 8 KubeHICE- Performance-aware Container Orchestration on Heterogeneous-ISA Architectures in Cloud-Edge Platforms
\cite{yang_kubehice_2021} & H/V    & \xmark & \cmark & \xmark & \xmark & S/A & \xmark & \xmark & \xmark      \\
% 9 BECA- A Blockchain-Based Edge Computing Architecture for Internet of Things Systems
\cite{ajayi_beca_2021-1} & H & \xmark & \xmark & \xmark & \cmark & A & \cmark & \xmark & \xmark      \\
% 10 Centralized and Federated Learning for Predictive VNF Autoscaling in Multi-domain 5G Networks and Beyond
\cite{subramanya2021centralized} & H/V & \cmark & \xmark & \cmark & \cmark & S/A & \xmark & \xmark & \xmark      \\
% 11 REACT- Enabling Real-Time Container Orchestration
\cite{struhar_react_2021} & \xmark & \xmark & \cmark & \xmark & \xmark & S & \xmark & \xmark & \xmark      \\
% 12 FLEDGE- Kubernetes Compatible Container Orchestration on Low-Resource Edge Devices
\cite{goethals_fledge_2020} & \xmark & \xmark & \cmark & \cmark & \xmark & \xmark & \xmark & \xmark & \xmark      \\
% 13 Distributed and Decentralized Orchestration of Containers on Edge Clouds
\cite{pires_distributed_2021} & \xmark& \xmark & \cmark & \xmark & \cmark & \xmark & \xmark & \xmark & \cmark      \\
% 14 A Blockchain-Based Containerized Edge Computing Platform for the Internet of Vehicles
\cite{cui_blockchain-based_2021} & \xmark& \xmark & \cmark & \xmark & \xmark & \xmark & \cmark & \xmark & \xmark      \\
% 15 A Decentralized Framework for Serverless Edge Computing in the Internet of Things
\cite{cicconetti_decentralized_2021} & \xmark & \xmark & \xmark & \cmark & \cmark & \xmark & \xmark & \xmark & \cmark      \\
% this work
This work & V    & \cmark & \cmark & \cmark & \cmark & S & \cmark & \cmark & \cmark \\
\hline
\end{tabular}
\label{table:literature_comparison}
\begin{tablenotes}[para]
\item [1] Description of Abbreviations: C1: Auto-scaling, C2: Predictive Scaling, C3: Software Heterogeneity, C4: Resource Constrained Hardware, C5: On-Device Orchestration, C6: Optimization Metrics, C7: Event-driven Communication, C8: Decentralized Application Delivery, C9: Decentralized Load Balancing, H: Horizontal, V: Vertical, A: Application, S: System
\end{tablenotes}
\end{threeparttable}
\end{table}

%todo cloudda ve edgede yapilan calismalara referans ver.

\section{Framework Architecture}

\begin{figure}
	\begin{center}
		\includegraphics[width=0.9\columnwidth]{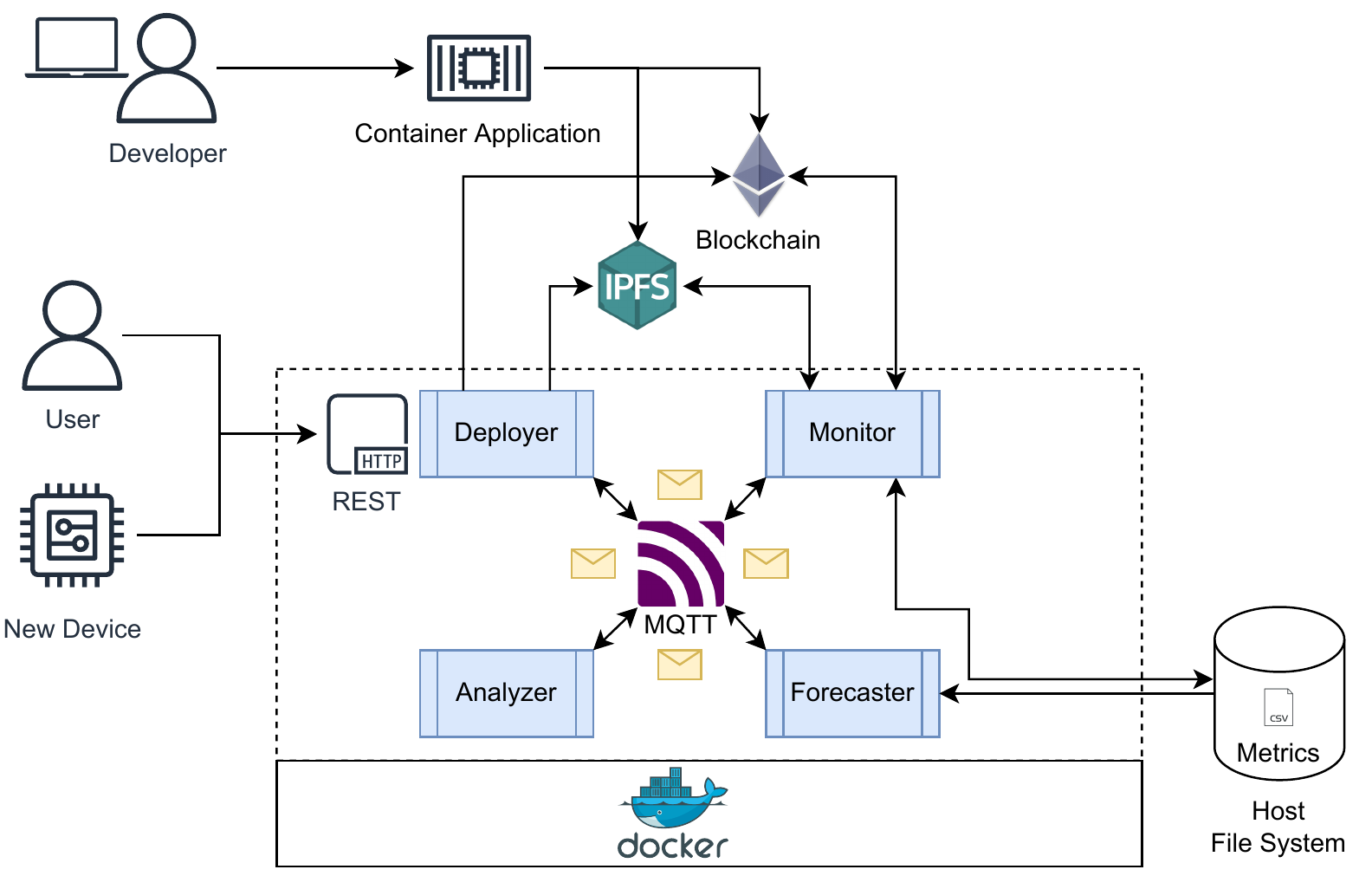}
		\vskip\baselineskip % Leave a vertical skip below the figure
		\caption{Overview of the framework's architecture on the edge.}
		\label{fig:architecture_orchestration}
	\end{center}
\end{figure}

The proposed framework manages the lifecycle of containerized applications, including design, delivery, execution, and optimization steps. It provides data storage, resource management, monitoring, application registry and delivery solutions, including service upgrades and software updates. The overview of the framework's architecture on the edge is illustrated in Figure \ref{fig:architecture_orchestration}. The framework on the edge is designed as a distributed set of components to support decentralized orchestration. It consists of four components deployed on the edge device. \textit{Monitor} scrapes metrics from the host system and shares them with the rest of the framework. \textit{Deployer} responds to application deployment and resource optimization requests. New deployment requests are accepted through a REST endpoint. \textit{Analyzer} evaluates incoming requests with the data made available by the other components and makes predictive analysis. \textit{Forecaster} provides time-series forecasting capabilities to the framework. 

The four components of the framework rely on an array of tools and technologies to form the full framework as shown in Figure \ref{fig:architecture_orchestration}. \textit{Docker} provides an engine with virtualization capabilities to the edge and means to execute and monitor containerized applications. \textit{IPFS} grants the distributed file storage for the Docker registry and long-term storage for application metrics. \textit{IPFS-Backed Docker Registry (IPDR)} is a Docker registry proxy that utilizes IPFS for decentralized image storage \cite{noauthor_ipdr_2022}. \textit{Blockchain} hosts smart contracts that connect the application delivery process with the edge framework. \textit{Host File System} is a mounted volume through Docker to grant short-term storage for application metrics on the host device. The framework's components communicate with the publish/subscribe pattern. \textit{Message Queueing Telemetry Transport (MQTT) Broker}   is a prominent lightweight protocol for event-driven architectures on the edge with low resource and power consumption \cite{koziolek_comparison_2020}. Here, components publish messages on specific topics to pass events and data for subscribed components to pick up. Messages contains the \texttt{action} field for identification of a message's purpose. Subscribed components act upon the received \textit{action} and payload of a message. Table \ref{table:message_action_table} presents a list of supported actions and their target topics.

\begin{table}
\caption[Supported Orchestration Actions and MQTT Topics]{Supported Orchestration Actions and MQTT Topics}
\begin{center}
\begin{tabular}{|l|l|} \hline
\textbf{Action} & \textbf{Topic}\\\hline
\textbf{Deployment Request} & Deploy \\\hline
\textbf{Deployment \mbox{Analysis Request}} & Analyze \\\hline
\textbf{Deployment \mbox{Optimization Request}} & Analyze \\\hline
\textbf{Forecast Request} & Forecast \\\hline
\textbf{Forecast Response} & Forecast \\\hline
\textbf{Deployment Accept} & Deploy \\\hline
\textbf{Deployment Cancel} & Deploy \\\hline
\textbf{Deployment Update} & Deploy\\\hline
\textbf{Monitoring Result} & Monitor\\\hline
\end{tabular}
\label{table:message_action_table}
\end{center}
\end{table}

\subsection{Application Delivery}

\begin{figure*}[!t]
    \centering
    \includegraphics[width=\textwidth]{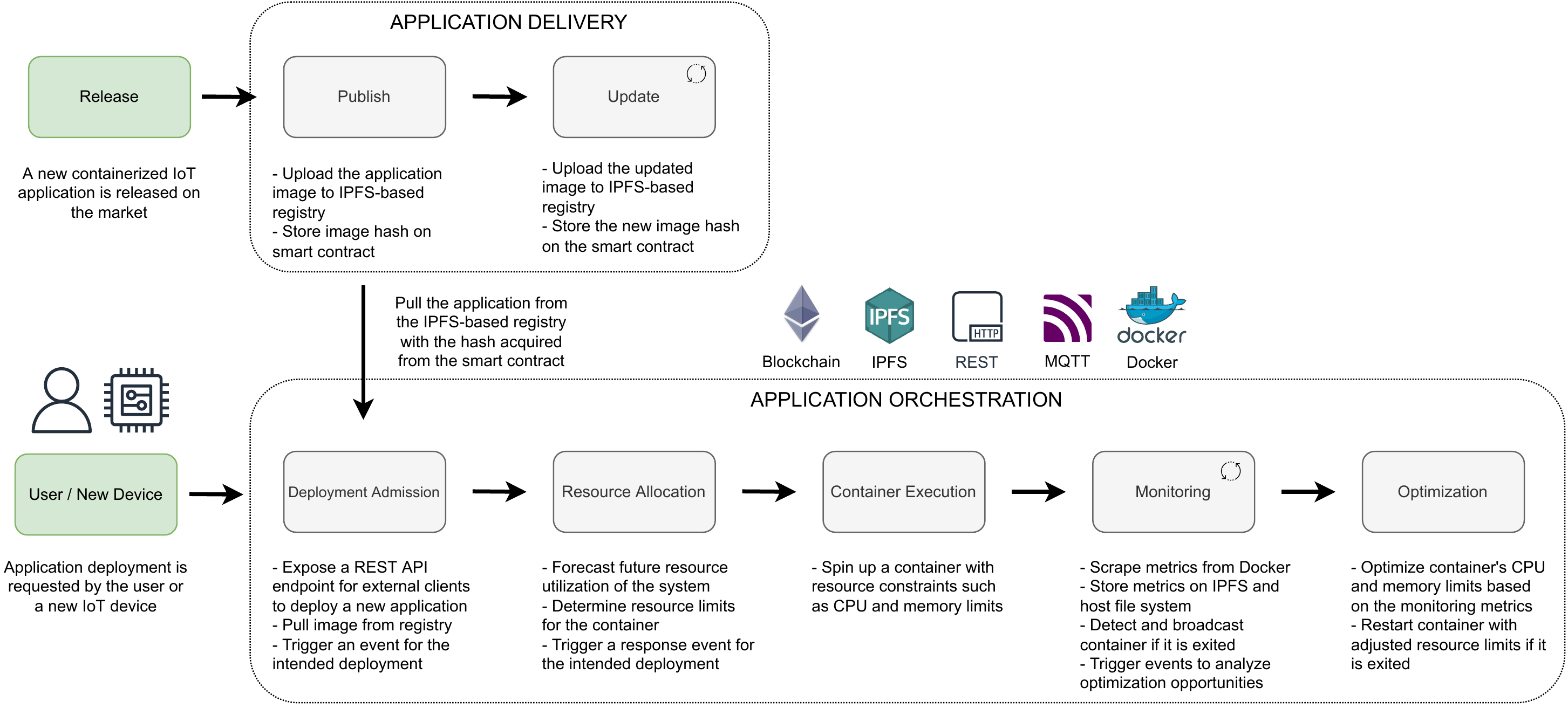}
    \caption{Application delivery and orchestration workflows of the framework.}
    \label{fig:architecture_framework}
\end{figure*}

The application delivery process provides a pipeline for developers to publish and deliver applications to their clients. The pipeline shown in Figure \ref{fig:architecture_framework} is built over a decentralized application registry. Versioning, updates, and upgrades are managed on top of IPFS using IPDR as a decentralized and agile solution. IPDR offers the advantage of downloading layers of a Docker image individually with the use of less bandwidth and local storage. In this work, an interface for this decentralized Docker registry is developed as a decentralized application with a smart contract using the data structures and functions in Listing \ref{lst:solidity}. Developers start the \textit{Release} process of a container application. The application release consists of either \textit{Publish}, or subsequent \textit{Update} processes recurring for each service upgrade and software update. First, the released or updated Docker image must be stored on IPDR, which returns an IPFS hash. Then, the application metadata stored on the blockchain is manipulated through the smart contract. The metadata consists of the image hash acquired from IPFS and an optional set of resource limits such as {\tt requestedLimit} and {\tt baseLimit} that can be configured by the developer. Data stored on IPFS cannot be traced to its owner unless proof of ownership is deliberately included; nevertheless, the smart contract allows developers to prove the application's authenticity. The delivery process ensures that applications are tamper-proof and only the owner can deliver updates and upgrades.

\begin{lstlisting}[language=Solidity, caption={Core data structure and functions for application delivery.},tabsize=2,frame=single,linewidth=0.99\columnwidth,xleftmargin=0.08\columnwidth,label={lst:solidity}]
    /* Image ownership */
    mapping(address => Image) images;
    /* Image metadata */
    struct Image {
        string imageHash;
        string imageName;
        uint baseLimitMemory;
        uint requestLimitMemory;
        uint baseLimitCPU;
        uint requestLimitCPU;
    }
    /* Get image metadata */
    function get(address imageOwner, 
            string memory imageName) 
            public view returns (Image memory);
    /* Set image metadata */
    function set(string memory imageHash,
                    string memory imageName, 
                    uint baseLimitMemory,
                    uint requestLimitMemory,
                    uint baseLimitCPU,
                    uint requestLimitCPU
                    ) public;
\end{lstlisting}

\subsection{Application Orchestration}

The framework components shown in Figure \ref{fig:architecture_orchestration} asynchronously pass messages between each other to provide dynamic orchestration shown in Figure \ref{fig:architecture_framework}. The messaging architecture is inspired by the \textit{service providers} and \textit{session messages} described in \cite{nalin_orchestration_2013}. The orchestration flow is a step-by-step process executed on the edge.  It is responsible for managing incoming application deployments and dynamically adapting allocated resources for each container. The self-adaptive properties are modeled after Monitor-Analyze-Plan-Execute-Knowledge (MAPE-K) loops \cite{arcaini_modeling_2015}. 

\subsubsection{Deployment Admission}

The deployment request is initiated by a request that is made externally using the REST API provided by the \textit{Deployer}. %todo How is the user/device aware of this service? Is user/device also registered to the sc?
The actor invoking the request can either be the user or an IoT device connected to the network. Then, the \textit{Deployer} acquires the image hash and resource limits stored on the blockchain via the smart contract. After pulling the corresponding application image from the registry, a \textit{deployment analysis request} is published with the resource limits on the \textit{analysis} topic.

\subsubsection{Resource Allocation}
\label{chap:resource_allocation_step}

Each container is deployed with a set of CPU and memory limits specified in the smart contract. These limits are used in scaling the containers based on the status of other active deployments and vendor-defined resource limits. The \textit{Analyzer} is subscribed to two different topics: \textit{analyze} and \textit{forecast}. It listens to \textit{deployment analysis requests} and publishes a corresponding \textit{deployment analysis response} on the \textit{analyze} topic as shown in Table \ref{table:message_action_table}. The analysis determines the feasibility of a new deployment with its resource requirements. 

The analysis takes the future availability of the resources into account rather than their current values to ensure the long-lasting stability of the system. Therefore, a time-series forecasting analysis is conducted by the \textit{Forecaster}. It is subscribed to the \textit{forecast} topic and acts on a \textit{forecast request} messages published by the \textit{Analyzer}. The Autoregressive Integrated Moving Average (ARIMA) model is used to interpret the trends in time-series data and predict the future values with statistical analysis \cite{noauthor_time_nodate, noauthor_statsmodelstsaarimamodelarima_nodate}. The time-series data consist of the metrics collected and stored locally by the \textit{Monitor} as described in Section \ref{chap:monitoring_step}. \textit{Auto-Regression} and \textit{Integrated} models of ARIMA are configured together to capture non-stationary patterns seen in each analyzed metric and forecast their values. The ARIMA model used in our implementation is with an order of (5,1,0), which can capture the short time trends in available metrics data. This configuration sets up a lag order of 5 to smooth the time-series data and a degree of differencing of 1. The use case scenarios depend on the environment where the edge unit is deployed. This work focuses on daily patterns which can be observable in smart homes or offices. In order to take this temporal characteristic into account, data points are aggregated hourly. %Then, the maximum value observed in the same or adjacent hours’ aggregate is taken. This value is then compared with each prediction result. Any prediction lower than the maximum observed value is replaced with the maximum observed value. 

\begin{comment}
represented as

\begin{equation}
\label{eqn:arima}
\begin{gathered}
y^{\prime}_{t} = c 
    + \phi_{1} y^{\prime}_{t-1} 
    + \phi_{2} y^{\prime}_{t-2} 
    + \phi_{3} y^{\prime}_{t-3} 
    + \phi_{4} y^{\prime}_{t-4} 
    + \phi_{5} y^{\prime}_{t-5}
\end{gathered}
\end{equation}
where $y^{\prime}_{t} = y_{t} - y_{t-1}$. 
\end{comment}

The prediction results for each container are its predicted CPU and memory utilization based on its lifecycle. These results, $P^{util}_{r,c}$, are published on the \textit{forecast} topics so that \textit{Analyzer} will be able to compute the feasibility of deploying a new container. To achieve this, it has to compute system availability, $P^{avail}_{r,S}$. Existing containers' resources are always prioritized over a new deployment. Hence, their resources,  $\mathbf{L}^{current}_{r,c}$, are not lowered. Therefore, $\mathbf{P}^{util}_{r,c}$ is updated as  $\mathbf{P}^{util}_{r,c}[t] := max(\mathbf{L}^{current}_{r,c}, \mathbf{P}^{util}_{r,c}[t]$) for each time point $t$. Predicted availability of each resource in the host $\mathbf{P}^{avail}_{r,S} := S^{total}_{r}$ is updated as $\mathbf{P}^{avail}_{r,S} \minuseq max_t(\mathbf{P}^{util}_{r,c}[t])$ for each $C_c$ and $R_r$. The symbol list is given in Table  \ref{table:table_of_symbols}.

\begin{table}
\caption[Table of Symbols.]{Table of Symbols.}
\begin{center}
\begin{tabular}{|l|l|} \hline
\textbf{Symbol} & \textbf{Description}\\\hline
\textbf{$R_r$} & Resource $r \in \{cpu,mem\}$ \\\hline
\textbf{$C_c$} & Container $c$ \\\hline
\textbf{$S$} & System \\\hline
\textbf{$S^{total}_{r}$} & Total resources $R_r$ of the system S \\\hline
\textbf{$S^{avail}_r$} & Availability of resource $R_r$ of the system S \\\hline
\textbf{$C^{util}_{r,c}$} & Utilization of resource $R_r$ by container $C_c$ \\\hline
$P^{avail}_{r,S}$ & {Predicted availability of resource $R_r$ of the system S} \\\hline
\textbf{$P^{util}_{r,c}$} & Predicted utilization of resource $R_r$ by container $C_c$\\\hline
\textbf{$P^{throttle}_{c}$} & Predicted CPU throttling percentage of container $C_c$ \\\hline
\textbf{$L^{current}_{r,c}$} & Current limit definition of resource $R_r$ for container $C_c$ \\\hline
\textbf{$L^{target}_{r,c}$} & Target limit definition of resource $R_r$ for container $C_c$ \\\hline
\textbf{$L^{request}_{r,c}$} & Request limit definition of resource $R_r$ for container $C_c$ \\\hline
\textbf{$L^{base}_{r,c}$} & Base limit definition of resource $R_r$ for container $C_c$ \\\hline
$L^{throttle}$ & Limit CPU throttling percentage \\\hline
\textbf{$L^{buffer}_{r}$} & Buffer ratio for scaling of resource $R_r$ \\\hline
\textbf{$L^{scale}_{r,i}$} & Limit scaling amount of $R_r$ for $i \in \{up,down\}$ \\\hline
\end{tabular}
\label{table:table_of_symbols}
\end{center}
\end{table}

The analysis request includes $L^{target}_{r,c}$ which can be one of the resource limit values: $L^{request}_{r,c}$, or $L^{base}_{r,c}$. In their absence, a default value is used. The \textit{Analyzer} approves a deployment request if $L^{target}_{r,c} \leq P^{avail}_{r,S}$, meaning that the system will have enough resources to accommodate the incoming application deployment. Otherwise, the request is rejected.

\subsubsection{Container Execution}
\label{chap:container_execution_step}

Deployment analysis responses received by the \textit{Deployer} can be twofold, corresponding to either a \textit{deployment accept} or a \textit{deployment cancel} as shown in Table \ref{table:message_action_table}. If a deployment is tried with $L^{request}_{r,c}$ and rejected, the \textit{Deployer} sends a second analysis request with $L^{base}_{r,c}$. For accepted deployments, a container is executed with $L^{target}_{r,c}$, successfully allocating the required amount of resources.

\subsubsection{Monitoring}
\label{chap:monitoring_step}

Resource utilization metrics are acquired from Docker API by calling the relevant endpoints and \textit{cgroup} stats in Linux systems by the \textit{Monitor}. These metrics are published as a \textit{Monitoring Result} on the \textit{monitor} topic and also stored locally. After the retention time configured in the framework is reached, the metrics are uploaded to IPFS, and their hashes are stored on the blockchain via a smart contract as long-term storage. \textit{Monitor} also identifies containers that have stopped prematurely. New \textit{deployment request} messages are published based on the retry counts. The \textit{Monitor} also initiates the optimization process for active containers.

\subsubsection{Optimization}
\label{chap:optimization_step}

In this step, the framework dynamically adjusts the resource limits of active containers. The aim is to downscale underutilized containers and upscale containers approaching the limits. \textit{Analyzer} determines whether the optimization is necessary. The target resource limits, $L^{target}_{r,c}$ are calculated based on the predictions provided by the \textit{Forecaster}. Then, an analysis response message is published on the \textit{deploy} topic for the container if it is scheduled for resource optimization. Finally, the \textit{Deployer} interacts with the Docker engine if there is a need for optimization. 

Optimization analysis starts with the calculation of $P^{avail}_{r,S}$, identical to the process explained in Section \ref{chap:resource_allocation_step}. Then, $L^{target}_{r,c}$ is derived from $L^{current}_{r,c}$ for CPU and memory, respectively as explained in the following paragraphs. After every approved optimization response, $P^{avail}_{r,S}$ is updated with $\mathbf{P}^{avail}_{R,S} := \mathbf{P}^{avail}_{R,S} - \Delta \mathbf{L}^{R,c}$ where $\Delta \mathbf{L}^{r,c} := \mathbf{L}^{target}_{r,c} -\mathbf{L}^{current}_{r,c}$.

Optimization for memory limit sets $L^{target}_{mem,c}$, which is bounded by system-defined minimum and maximum resource values, $L^{min}_{mem}$ and $L^{max}_{mem}$. The upper limit ensures that a single container cannot allocate a very high share of system memory. The lower limit protects the container from scaling down indefinitely. The amount for scaling up and down,  $L^{scale}_{mem,up}$ and $L^{scale}_{mem,down}$, are constants that can be configured in the framework. However, since $C^{util}_{mem,c}$ exceeding $L^{current}_{mem,c}$ is a risk that can kill the container, a margin is defined between $P^{util}_{mem,c}$ and $L^{target}_{mem,c}$ while scaling up. If $P^{util}_{mem,c}$ is less than $C^{util}_{mem,c}$, the \textit{Analyzer} computes $\mathbf{L}^{target}_{mem,c} $ as $ \mathbf{L}^{current}_{mem,c} - \mathbf{L}^{scale}_{mem,down}$ to scale down the container. After the $L^{target}_{mem,c}$ is set, in case of downscaling, the new limit is once again compared with $max(\mathbf{P}^{util}_{mem,c})$ to make sure that the container is not scaled below the allocated margin. Otherwise, $L^{current}_{mem,c}$ is returned, and memory is not scaled down any further. For scaling up, a similar approach is adopted.

Optimization for CPU limit sets $L^{target}_{cpu,c}$. If $max(P^{util}_{cpu,c})$ is greater than $L^{current}_{cpu,c}$, container is upscaled with $\mathbf{L}^{target}_{cpu,c} := \mathbf{L}^{current}_{cpu,c} + \mathbf{L}^{scale}_{cpu,up}$. In the opposite case, container is downscaled with $\mathbf{L}^{target}_{cpu,c} := \mathbf{L}^{current}_{cpu,c} - \mathbf{L}^{scale}_{cpu,down}$. Setting a limit on the CPU causes throttling in Docker \cite{noauthor_runtime_2022}. As high amount of $C^{throttle}_{c}$ severely degrades the application performance, two extra precautions are taken while downscaling the containers that have already been throttling. Firstly, $P^{throttle}_{c}$ is compared against $L^{throttle}$ which is a constant defined in our framework. If $max_t(P^{throttle}_{c}[t])$ is greater than $L^{throttle}$, an \textit{adjustedScale} amount where $adjustedScale := (\mathbf{L}^{scale}_{cpu,up} \times max_t(P^{throttle}_{c}[t])) / 100$ is used instead of $\mathbf{L}^{scale}_{cpu,up}$ for a minor scale up. The second control is applied before returning $L^{target}_{cpu,c}$, which can cause throttling after a scale-down. In order to prevent sudden throttling, $\mathbf{L}^{target}_{cpu,c}$ is recalculated as $\mathbf{L}^{target}_{cpu,c} := max(P^{util}_{cpu,c}) \times \mathbf{L}^{buffer}_{cpu}$, which is set to 110\% in our framework. Buffer ensures the application to have enough room for unexpected load spikes until the next optimization without causing unnecessary throttling. After this adjustment, it is possible that $L^{target}_{cpu,c}$ can be above $L^{current}_{cpu,c}$. In that case, any scaling operations are overturned by returning $L^{current}_{r,c}$ as the throttling was already below $L^{throttle}$.

\subsection{Cluster Deployment}
\label{chap:cluster_deployment}

The proposed framework can be deployed on multiple edge devices as a cluster in the same network. This strategy supports not only load balancing across the connected hosts but also decentralized orchestration of deployments. The setup depends on MQTT bridges where brokers can automatically broadcast configured messages in the network. Each device's IP address is provided to each other broker to set up the bridges. However, it is also possible to set up an MQTT device discovery mechanism to automatically find other brokers on the network \cite{rende_dag_2017}. Shared topics, \textit{monitor} and \textit{deploy}, are configured and prefixed with the \textit{cluster} keyword so that each component can differentiate internally or externally generated messages.

\textit{Deployer} on each device is subscribed to both \textit{monitor} and \textit{cluster/monitor} topics. Key/value pairs of IP addresses and $S^{avail}_r$ are generated by the \textit{Deployer} for each device and updated with each \textit{monitoring result} message. \textit{Deployer} uses these pairs to track the available resources of all devices in the network. It is also subscribed to the \textit{deploy} and \textit{cluster/deploy} topics where it receives deployment requests made on all devices. When a new request is received, each \textit{Deployer} selects the device with the highest amount of available resources from its key/value pairs. Then, only the device that decides itself as the best candidate continues with the deployment by publishing a \textit{deployment analysis request}. Cluster deployment follows the same steps as the regular deployment workflow. If $S^{avail}_r$ of multiple devices are the same, each device marks the device with the smallest IP address for deployment. 

\section{Experiments and Results}
\label{chap:experiments_and_results}

All experiments are carried out on Raspberry Pi 4 Model B with a 64-bit quad-core Cortex-A72 processing unit,  8GB LPDDR4-3200 SDRAM, and Broadcom BCM2711. A slice with 1 CPU and 1GB memory with disabled swap usage is created to emulate a resource-constrained device. Docker daemon's \textit{cgroup} parent is assigned to this slice to limit the maximum allowed resource usage. CPU limits and utilization are given in CPU units. 1 CPU is equal to a single CPU core, and 50\% of this core is represented as 500mCPU. The framework also exists on the same hardware. The CPU utilization is negligible for most components except for the \textit{Forecaster} and IPFS. The \textit{Forecaster} utilizes all available CPUs during forecasting to produce results as quickly as possible. IPFS utilizes 50-250 mCPU on average. It can be reduced by switching to a minimal setup, disabling unnecessary features, and limiting the peers. The memory footprint of the entire framework is currently ~400MB, which can be reduced significantly by pruning and optimizing the libraries.

%All experiments are carried out on Raspberry Pi 4 Model B with a 64-bit quad-core Cortex-A72 processing unit,  8GB LPDDR4-3200 SDRAM, and Broadcom BCM2711. A slice with 1 CPU and 1GB memory with disabled swap usage is created to emulate a resource-constrained device. Docker daemon's \textit{cgroup} parent is assigned to this slice to limit the maximum allowed resource usage. CPU limits and utilization are given in CPU units. 1 CPU is equal to a single CPU core, and 50\% of this core is represented as 500mCPU. The framework also exists on the same hardware. The CPU utilization is negligible for most components except for the \textit{Forecaster} and IPFS. The \textit{Forecaster} utilizes all available CPUs during forecasting to produce results as quickly as possible. IPFS utilizes 50-250 mCPU on average. It can be significantly reduced by switching to a minimal setup, disabling unnecessary features, and limiting the number of connected peers. The memory footprint of the entire framework is currently ~400MB, which can be reduced significantly by pruning and optimizing the libraries.

\begin{figure*}
	\begin{center}
		\includegraphics[width=2\columnwidth]{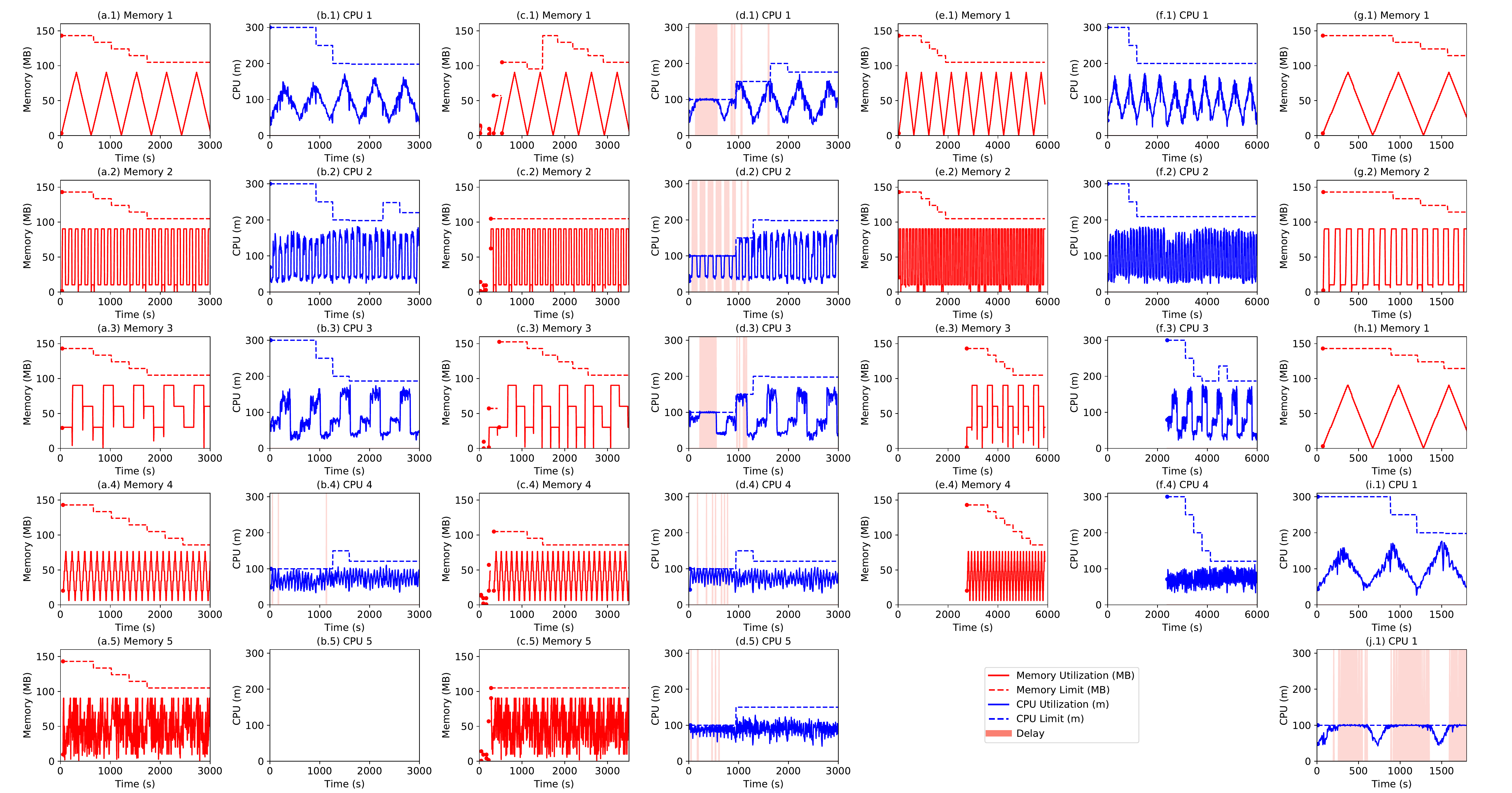}
		\vskip\baselineskip % Leave a vertical skip below the figure
		\caption[Workload CPU/memory utilization/limit and delays observed in each experiment. ]
    {\tabular[t]{@{}l@{}}Workload CPU/memory utilization/limit and delays observed in each experiment. \\ 
		 Sequential deployments with ample resource specification: Memory (a), CPU (b).
		 \\
		 All workloads with drastically low resource specification: Memory (c), CPU (d).
		 \\
		 Interleaved deployment of workload pairs: Memory (e), CPU (f).
		 \\
		 Extreme resource constraints on the system: Memory (g,h), CPU (i,j). \endtabular}
		\label{fig:graph}
	\end{center}
\end{figure*}

In order to create a controlled environment for the tests, containerized applications are prepared for each resource type. Five workload patterns proposed in \cite{taherizadeh2019dynamic} for container-based cloud applications are adapted to this work to generate workload patterns that represent different IoT application types:

\begin{enumerate}
    \item Slowly rising/falling workload pattern
    \item Drastically changing workload pattern
    \item On-off workload pattern
    \item Gently shaking workload pattern
    \item Real-world workload pattern
\end{enumerate}

The IoT applications can be classified as data-dominant and CPU-dominant. Hence, ten containerized applications are generated. Labels for each workload pattern are referenced with the class name. For example, a slowly rising/falling pattern for the data-dominant workload is identified as \texttt{Memory 1}. The same pattern for CPU-dominant workload is represented by \texttt{CPU 1}. The experiments are carried out on each group separately. This approach does not exclude the case where a CPU-dominant application coexists with a data-dominant application on the same device as different algorithms execute for memory optimization and CPU optimization. The maximum memory amounts in these workloads are 95MB, 95MB, 95MB, 80MB, and 95MB, respectively, as listed above. Similarly, 150m, 150m, 150m, 120m, and 140m represent the maximum CPU utilization. The experiments are carried out on each group separately. This approach does not exclude the case where a CPU-dominant application coexists with a data-dominant application on the same device as different algorithms execute for memory optimization and CPU optimization. The experimental results are presented in Figure \ref{fig:graph} where solid and dashed lines show workloads and resource limits, respectively. 

\subsection{Orchestration of Multiple Containers on a Single Device}

These experiments are carried on for a single device that has to host multiple containers with different limit settings and deployment scenarios. The first two scenarios consider application deployments in an extremely short period. This usually happens when the device is installed for the first time. The third scenario simulate deploying new applications on an already running device. The fourth scenario handles the case where the host is extremely low on resources
%In each experiment, a subset of containerized applications is deployed on the same device to demonstrate four different scenarios. % of co-located workloads.

\subsubsection{Sequential Deployment Requests with Ample Resource Limits}
\label{chap:exp1}
In this scenario, the application limits specified are above the expected peak memory usage of the workloads. For each data-dominant workload, $L^{request}_{mem,c}=150MB$ and $L^{base}_{mem,c}=100MB$. As shown in Figure \ref{fig:graph}(a.1-5), all five workloads start simultaneously. The memory limits converge to the actual usage in a few iterations of optimization steps every 5 minutes. The initial delay before the first optimization allows the framework to collect enough metrics to make utilization predictions for the upcoming optimization interval.

In CPU-dominant workloads, $L^{request}_{cpu,c}=300m$ and $L^{base}_{r,c}=100m$. Recall that if neither $L^{request}_{cpu,c}$ nor $L^{base}_{cpu,c}$ is accepted, the deployment is rejected because of the limited resource of the system. Note that total request makes 1500mcore which is more than the allocated CPU on the emulation hardware. Thus, the first deployment of the fourth workload with $L^{request}_{cpu,c}$ is rejected and deployed with $L^{base}_{cpu,c}$ in a second attempt by the framework, as shown in Figure \ref{fig:graph}(b.1-4). With the first four workloads running on the system, neither $L^{request}_{cpu,c}$ nor $L^{base}_{cpu,c}$ for the fifth workload can be satisfied (Figure \ref{fig:graph}(b.5)). The limits of the successfully deployed workloads are initially lowered drastically with a constant $L^{scale}_{cpu,down}$, which is later adjusted based on $P^{util}_{cpu,c}$ and {$P^{throttle}_{c}$} resulting in more minor decrements. %After the fourth optimization attempt, all limits converge to an upper limit of 10\% above the  $P^{util}_{cpu,c}$.

\subsubsection{Sequential Deployment Requests with Drastically Low Resource Specifications}
\label{chap:exp2}

This scenario assumes that all limits are given much lower than the actual requirements. In this experiment, we set $L^{request}_{mem,c}=15MB$ and $L^{base}_{mem,c}=10MB$ . All five memory workloads fail to start for the first couple of tries, as shown in Figure \ref{fig:graph}(c1-5). After the initial two deployment attempts with no prior knowledge of the application, the framework begins assigning by rapidly increasing $L^{target}_{mem,c}$ based on $L^{base}_{mem,c}$ incremented by $L^{scale}_{mem,c}$ by several retry counts. Then, after the first four tries, Memory 1 and Memory 2 can be deployed as their resource usage is comparatively lower than the rest in the first one-third of their period. Other workloads keep getting killed as they immediately try to allocate more memory than allowed by $L^{current}_{mem,c}$. These burst restarts cause their $L^{target}_{mem,c}$ to increase so rapidly that this limit surpasses  $C^{util}_{mem,c}$. However, once the containers are deployed, their limits are lowered by $L^{scale}_{down,mem}$ until a tighter fit based on $P^{util}_{mem,S}$ is found.

The deployment behavior of the CPU workloads vastly differs from the memory workloads. This is due to the absence of an error similar to \textit{out of memory}. The limits are set as $L^{request}_{cpu,c}=100m$ and $L^{base}_{cpu,c}=50m$. All five deployments are accepted simultaneously due to their misconfigured $L^{request}_{cpu,c}$ values. These workloads typically exhibit $C^{util}_{r,c}$ higher than  $L^{current}_{cpu,c}$ . Consequently, they are fully throttled and containers exhibit delays in their workload which is shown with light red vertical regions in Figure \ref{fig:graph}(d.1-5). The framework optimizes $L^{current}_{cpu,c}$ after a couple of optimization cycles, eventually matching the limits similar to those in Figure \ref{fig:graph}(b.1-5). Note that the delay is removed.

The deployment behavior of the CPU workloads vastly differs from the memory workloads, as shown in Figure \ref{fig:graph}(c.1-5). This is tied to the absence of an error similar to \textit{out of memory}. Therefore, all five deployments are deployed with their misconfigured $L^{request}_{cpu,c}$ values. These workloads typically exhibit a $C^{util}_{r,c}$ higher than this  $L^{current}_{cpu,c}$ . Consequently, they are all immediately throttled. Fully throttled containers exhibit delays in their workload which is demonstrated in Figure \ref{fig:graph}(d.1-5) for workloads \textit{CPU 1-5}. The framework optimizes $L^{current}_{cpu,c}$ responsible for the throttling after a couple of optimization cycles, eventually matching the limits similar to those in Figure \ref{fig:graph}(b.1-5).

\subsubsection{Interleaved Deployment of Workload Pairs}
\label{chap:exp3}

This experiment studies the effects of introducing a new container to a stabilized system. For this purpose, after the first two workloads are optimized, deployment requests from the last two workloads will be received.

As shown in Figure \ref{fig:graph}(e.1-4) and Figure \ref{fig:graph}(f.1-4), the outcome is the same for CPU and memory workloads. The optimization steps adapt to the new workloads without interfering with the past deployments. %The frameworks’ optimization loops are independent of the arrival of the new deployment. 
The optimization of the newly introduced applications starts once enough data is collected for predictions to be performed. 

\subsubsection{Extreme Resource Constraints on the System}
\label{chap:exp4}

Some hosts can be running on extremely low resources. Pre-existing deployments can impose very tight constraints on the new requests. To simulate this environment for data-dominated applications, we set $S^{avail}_{mem}=400MB $ and studied with the first three memory workloads. The first two workloads are deployed with their $L^{request}_{mem,c}$ (Figure \ref{fig:graph}(g.1-2)). The third workload cannot be deployed as there is not enough memory for the third workload's $L^{request}_{mem,c}$ or $L^{base}_{mem,c}$. Another experiment is run with a similar configuration with $S^{avail}_{mem}=200MB$. As shown in Figure \ref{fig:graph}(h.1), only the first deployment is accepted as any subsequent deployments would require more memory than $S^{total}_{mem}$.

Similar results are obtained from the same experiments for CPU images. We set  $S^{avail}_{cpu}=350m$ and deployment requests for the first three workloads are sent. Only the first one is accepted with $L^{request}_{cpu,c}$ as shown in Figure \ref{fig:graph}(i.1). However, $S^{avail}_{cpu}$ is not enough to deploy the remaining two workloads. These workloads are tried twice with $L^{request}_{cpu,c}$ and $L^{base}_{cpu,c}$ before they are rejected. Then, the experiment is rerun with even tighter resources where $S^{avail}_{cpu}=100m$. A similar deployment pattern can be seen in in Figure \ref{fig:graph}(j.1). The only difference is that the first workload is deployed with $L^{base}_{cpu,c}$. This container exhibits delays at regular intervals; nevertheless, it cannot be scaled up due to lack of $S^{avail}_{cpu}$.

\subsection{Experiments with Cluster Deployment}

This experiment is devised to observe the load balancing capabilities of the framework across multiple devices. Three edge devices with identical resources, $S^{avail}_r$, are used for this purpose. Device IP addresses increase from device 1 to 3, where \textit{device 1} is the device with the smallest IP address. The first device receives sequential deployment requests from the first four memory workload patterns, i.e. Memory 1-4. %The expected behavior for each device is the self-decision t decide to load balance deployment requests in a decentralized manner.

Firstly the MQTT broker on each device starts, and the framework is deployed. 
Then \textit{device 1} receives Memory 1 workload. It deploys it on itself since its IP is the smallest. Following this deployment, all three devices synchronize their resource availability key/value pairs with \textit{monitoring\_result} messages published by \textit{device 1}. With the second request, each device identifies \textit{device 2} as the processor. The same steps are followed for the remaining two applications. The third application is deployed on \textit{device 3} which evens up the resource availability of all three devices. The final application is deployed on \textit{device 1} which has the smallest IP address. 
It should be noted that the cluster deployment does not introduce latency overhead during deployments. Only a single extra event is exchanged over the \textit{cluster/deploy} topic if another device satisfies the deployment request while the rest of the workflow stays the same.

\section{Conclusion}

In this paper, we have presented an end-to-end solution for a seamless user experience between application developers and end-users on the IoT edge. The exploitation of virtualization and decentralization technologies transformed heterogeneous and resource-constrained edge gateways into vendor-agnostic hosts for IoT applications. Our framework lowers the edge computing footprint by sharing hardware and maximizing resource utilization with autonomous scaling and load balancing in the presence of multiple devices. The event-driven resource manager establishes a fair and reliable platform for all involved parties based on analysis of running services and forecasts of future requirements. Concerns about the authenticity of deliveries such as releases and upgrades are ensured through smart contracts and distributed files system. Specific aspects of the framework can be effortlessly extended due to its modular architecture of message queues and distributed components. %Diversifying optimization analysis results with container migration will enhance the flexibility of orchestration and stability of deployments.

\section*{Acknowledgments}

This research is partially supported by Boğaziçi University Scientific Research Projects No: 17A01P7 and Textkernel B.V.

\bibliographystyle{IEEEtran}
\bibliography{IEEEabrv,references.bib}

\end{document}